\documentclass[12pt]{article}

\usepackage{amssymb}

\usepackage{lineno}

\usepackage{textgreek}

\usepackage[unicode=true,pdfborder={0 0 0 [0 0]},colorlinks=true,linkcolor=blue,citecolor=blue,urlcolor=blue]{hyperref}

\usepackage[utf8]{inputenc}
\usepackage[T1]{fontenc}

\usepackage{tabularx}

\usepackage{authblk}

\usepackage{graphicx}

\usepackage{subfigure}

\usepackage[labelfont=bf]{caption}

\usepackage{wrapfig}

\usepackage{indentfirst}

\usepackage[left=1.0in,right=1.0in,top=1.0in,bottom=1.0in,includeheadfoot]{geometry}

\usepackage{multirow}

\newcommand{\lead}{\textsuperscript{206}Pb}

\newcommand{\leadams}{\textsuperscript{210}Pb}
\newcommand{\polonium}{\textsuperscript{210}Po}
\newcommand{\leadle}{(\textsuperscript{206}PbF\textsubscript{3})\textsuperscript{-}}
\newcommand{\leadamsle}{(\textsuperscript{210}PbF\textsubscript{3})\textsuperscript{-}}
\newcommand{\leadhe}{\textsuperscript{206}Pb\textsuperscript{3+}}
\newcommand{\leadamshe}{\textsuperscript{210}Pb\textsuperscript{3+}}
\newcommand{\fluoride}{PbF\textsubscript{2}}
\newcommand{\hydroxide}{Pb(OH)\textsubscript{2}}

\newcommand{\customfootnotetext}[2]{{
  \renewcommand{\thefootnote}{#1}
  \footnotetext[0]{#2}}}
  
\begin{document}

\setlength{\parindent}{16pt}

\setlength{\parskip}{7pt}

\date{}

\title{\textbf{\leadams\ measurements at the André E. Lalonde AMS Laboratory for the radioassay of materials used in rare event search detectors}}

 \author[a,b,*]{Carlos~Vivo-Vilches}

 \author[a,b]{Benjamin~Weiser}

 \author[c,d]{Xiaolei~Zhao}
 
 \author[c]{Barbara~B.A.~Francisco}
 
 \author[a]{Razvan~Gornea}

 \author[c,d]{William~E.~Kieser}

 \affil[a]{Department of Physics, Carleton University, 1125~Colonel~By~Drive, Ottawa, ON~K1S~5B6, Canada}
 \affil[b]{Arthur B. McDonald Institute, 64~Bader~Lane, Queen's University, Kingston, ON~K7L~3N6, Canada}
 \affil[c]{A. E. Lalonde AMS Laboratory, University of Ottawa, 25~Templeton~St., Ottawa, ON~K1N~6N5, Canada}
 \affil[d]{Department of Physics, University of Ottawa, 25~Templeton~St., Ottawa, ON~K1N~6N5, Canada}
 
\maketitle

\customfootnotetext{*}{Corresponding author at Department of Physics, Carleton University, 1125~Colonel~By~Drive, Ottawa, ON~K1S~5B6, Canada\\
\emph{E-mail:} \href{mailto:cvivo@physics.carleton.ca}{cvivo@physics.carleton.ca} (C. Vivo-Vilches)}

\rule{\textwidth}{1pt}

\begin{abstract}
Naturally occurring radionuclide \leadams\ ($T_{1/2}$=22.2~y) is an important source of background in rare event searches, like neutrinoless double-\textbeta\ decay and dark matter direct detection experiments. \textgamma -counting measurements are performed when a sample mass of hundreds of grams is available. There are other cases, though, where only grams of sample can be used. In these cases, better sensitivities are required.

In this paper the capabilities of the A.E. Lalonde AMS Laboratory at the University of Ottawa for \leadams\ measurements are presented, in collaboration with the Astroparticle Physics group at Carleton University. \fluoride\ or PbO targets can be used, selecting in the low energy sector, respectively, (PbF\textsubscript{3})\textsuperscript{-} or (PbO\textsubscript{2})\textsuperscript{-} ions.

For fluoride targets, the blank \leadams /\lead\ ratio is in the 10\textsuperscript{-14}-10\textsuperscript{-13} range, but current output is lower and less stable. For oxide targets, current output shows better stability, despite a significant difference in current output of commercial PbO and processed samples, and background studies suggest a background not much higher than that of fluoride targets.

Measurements of Kapton films, an ultra-thin polymer material where masses available are typically just several grams, were performed. 90\%~C.L. upper limits for the \leadams\ specific activity in the range of 0.74-2.8~Bq/kg are established for several Kapton HN films.
\end{abstract}


\rule{\textwidth}{1pt}


\section{Introduction}
\label{sec:intro}

Radioassay of materials is one of the most important tasks in the design and construction of high sensitivity detectors for rare event searches, like direct dark matter detection \cite{NEWSG,PICO,DEAP2018,Lehnert2018}, and the observation of neutrinoless double-beta decay \cite{EXO-200,EXO-radioassay,Majorana2019,Majorana-radioassay,nEXO-PCR,nEXO2018-sensitivity}.

Assay of naturally occurring radionuclides producing decay chains, like \textsuperscript{238}U, is especially important, since each decay of the parent radionuclide would cause the production of several particles, each of which induce background for the experiment. These include \textgamma, \textbeta , and \textalpha\ particles. In secular equilibrium, all nuclides from the chain would have an activity similar to the parent radionuclide. It is quite common, however, that one of the \textsuperscript{238}U daughter radionuclides, \leadams, does not fulfill this condition.

In samples where all the radionuclides from the \textsuperscript{238}U decay chain are in secular equilibrium, the equal activity would lead to a \textsuperscript{238}U\ concentration 2~$\times$~10\textsuperscript{8} times higher than that of \leadams. Nevertheless, the secular equilibrium can be broken by the noble gas \textsuperscript{222}Rn ($T_{1/2}$=3.8235~d). This radionuclide outgasses from any material where \textsuperscript{238}U is present and, consequently, is present in air. The deposition and/or plating-out of \textsuperscript{222}Rn daughters from the air to materials which only have an ultra-trace concentration of uranium, potentially could increase concentrations of \leadams\ above the equilibrium with \textsuperscript{238}U. If this process takes place during the production of the material, it can even be present in the bulk of the final material, and not just on its surface. 

The main impact of \leadams\ in rare event searches comes from the \textalpha -decay of its daughter nuclide, \polonium . The decay itself can be a direct source of background in experiments looking for weakly interacting massive particles (WIMPs), one of the candidates for dark matter. In addition, these \textalpha\ particles can produce neutrons through (\textalpha ,n) reactions on low-Z elements. Because of the low range of \textalpha\ particles, this interaction typically takes place within the same material where the \textalpha -decay occurs.

Neutrons are also a direct source of background in WIMP searches, because they deposit energy through nuclear collisions which mimic the expected signals from these WIMPs \cite{STEIN201892}. In experiments searching for neutrinoless double-\textbeta\ decay, \textalpha\ particles are not a direct source of background, but neutrons produced by (\textalpha ,n) reactions activate materials, thereby increasing the background. 

Taking into account its relatively short half-life (22.2 y), \leadams\ is typically measured by radiometric techniques \cite{HOU2008105}:

\begin{itemize}
    \item Direct \textgamma-counting of the 46.539~keV \textgamma\ produced in 4.25\% of the \leadams\ decays. There are 2 challenges related to this measurement: the low intensity of this \textgamma\ decay; and the very low energy of the photon, for which the detection efficiency of germanium detectors is low.
    \item \textbeta-counting of the decay of the \leadams\ immediate daughter, \textsuperscript{210}Bi ($T_{1/2}=\mathrm{5.012~d}$). This method requires an extensive chemical preparation to separate \textsuperscript{210}Bi from other radionuclides, since most of them could cause a background in the \textbeta\ continuous spectrum. Because of the much shorter half-life of \textsuperscript{210}Bi, secular equilibrium with \leadams\ can be assumed.
    \item \textalpha-counting of the \polonium\ decay ($T_{1/2}=\mathrm{138.376~d}$). The main disadvantage of this technique is that secular equilibrium cannot be always assumed. If \leadams\ is deposited in the material only a few days before the measurement, \polonium\ activity will not be in equilibrium yet, and will be much lower. In other cases, the chemical production of the sample can lead to \polonium\ activities higher than \leadams\ ones, if the processes have a higher chemical efficiency for polonium. Direct \leadams\ measurements, therefore, will be more useful. \polonium\ \textalpha\ detection is useful and efficient for geological applications, where equilibrium can be assumed. When \leadams\ is measured by this technique, the original amount of \polonium\ in the sample has to be chemically removed. Then, several months are required for the \leadams\ to decay, leading to a measurable \polonium\ activity \cite{Ebaid}.
\end{itemize}

Regarding sensitive assays, these techniques require large amounts of sample, i.e. several hundreds of grams. For some materials, especially low density polymers, it would be an issue. For instance, a 30~cm~$\times$~30~cm film of 25.4~\textmu m thick Kapton weights only 3.25~g. 

ICP-MS allows rapid \leadams\ measurements, but with limited sensitivity, due to molecular background \cite{LARIVIERE2005188}. Therefore, for rapid and ultra-sensitive \leadams\ assay, accelerator mass spectrometry (AMS) is an alternative approach worth exploring. The possibility of employing AMS, using lead fluoride targets, to measure \leadams\ was first discussed about 20 years ago \cite{Steier2002,VOCKENHUBER2003713}. Most recently, the André E. Lalonde AMS Laboratory (AEL-AMS) has performed similar studies \cite{Sookdeo2015,Sookdeo2016}. These have corroborated the observation that the extraction of (PbF\textsubscript{3})\textsuperscript{-} ions from PbF\textsubscript{2} targets is a good choice, offering a good ionization efficiency \cite{KORSCHINEK1988328,Zhao2010}.

Bibliography shows that a relatively high and stable current of (PbO\textsubscript{2})\textsuperscript{-} ions can be extracted when sputtering PbO samples mixed with silver powder \cite{cookbook}. Nevertheless, this type of target had not been used before for \leadams\ AMS because of the injection of (\textsuperscript{208}Pb\textsuperscript{16}O\textsuperscript{18}O)\textsuperscript{-} ions into the accelerator when selecting the (\textsuperscript{210}Pb\textsuperscript{16}O\textsubscript{2})\textsuperscript{-} ion, potentially leading to a large, nearly isobaric, interference from \textsuperscript{208}Pb ions. Alternatively, fluorine has only one stable isotope, \textsuperscript{19}F, so the only (PbF\textsubscript{3})\textsuperscript{-} ion injected when selecting a mass of 267~u is \leadamsle. A post-accelerator spectrometer with high resolution, as in the case of the AEL-AMS system, sufficiently reduces any background from molecular fragments.

In this work, the use of this 3 MV AMS system for the \leadams\ assay in materials considered for the construction of low background detectors is evaluated, with particular attention to rare event searches at SNOLAB. The study of the performance parameters of \leadams\ measurements in the 3 MV AMS system, for the two different target materials, \fluoride\ or PbO, is presented in \autoref{sec:performance}, proving that AMS is one of the most appropriate techniques for \leadams\ assay. In \autoref{sec:application}, the interest of \leadams\ AMS measurements in polymer materials for rare event searches is addressed, as well as the chemical method and first results for Kapton polyimide film, showing specific activities below 1~Bq/kg using less than 2~g of sample.

\section{Performance of \textsuperscript{210}Pb measurements at the AEL-AMS facility}
\label{sec:performance}

\subsection{Measurement procedure}

\begin{figure}[t!]
    \centering
\includegraphics[width=\textwidth]{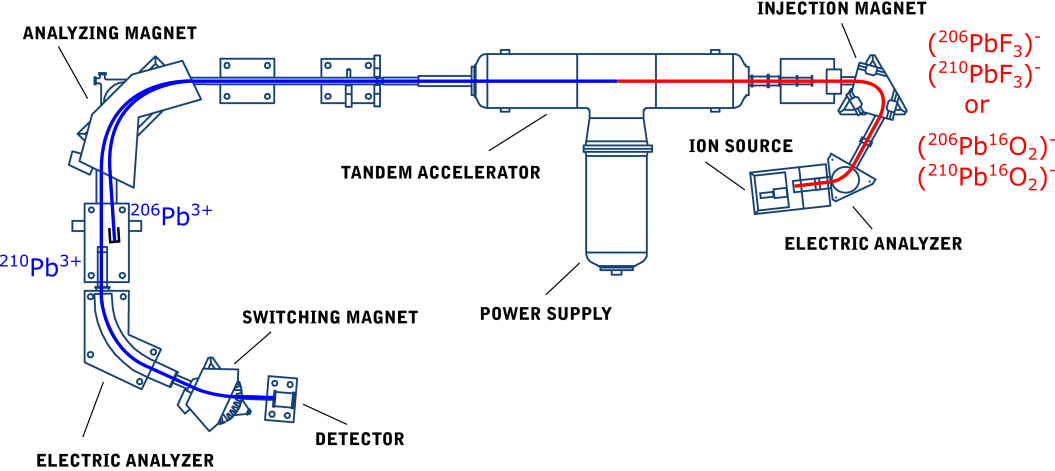}
    \caption{Illustration of the measurement set-up for \leadams\ measurements at the AEL-AMS system for each of the chemical species used.}
    \label{fig:leadams-measurement}
\end{figure}

All the measurements are performed with the 3 MV system of the AEL-AMS Laboratory at the University of Ottawa \cite{Kieser2015}. Illustrations of the set-up of this system for \leadams\ measurements, depending on the chemical species used, are presented in \autoref{fig:leadams-measurement}.

Samples are inserted as \fluoride\ or PbO, and the (PbF\textsubscript{3})\textsuperscript{-} or (PbO\textsubscript{2})\textsuperscript{-} ion, respectively, is selected by the low energy spectrometer. In both cases, the sample is mixed with silver powder in a volume ratio of 1:1. The ions are accelerated using a terminal potential of 2.5~MV in the tandetron accelerator. Ar gas stripper pressure is 0.012~mbar. With the high energy magnet, the charge state 3+ is selected; therefore, the final energy of the ions is 9.47~MeV in the case of measurements with \fluoride, and 9.67~MeV with PbO.

Following the high energy magnet, with a bending radius of 2~m, \leadhe\ ion current is measured with an off-axis Faraday cup (FC). This natural isotope of lead is chosen rather than \textsuperscript{208}Pb as the offset Faraday cup in the \textsuperscript{208}Pb position would block the \leadams\ beam.  This allows the use of the same terminal voltage for both the \lead\ and \leadams\ isotopes which permits the use of the more efficient fast sequential injection method. The \leadamshe\ ions continue along the beamline, through the electrostatic analyzer (ESA) and the switching magnet, until they reach the gas ionization chamber detector.

All the pre-accelerator slits are set to $\pm$1.25~mm, while all the post-accelerator slits, to $\pm$1.00~mm.

\subsection{Ionization efficiency and ion transmission}

Several parameters related to the \leadams\ measurements with AEL-AMS system are summarized in \autoref{tab:efficiency}. When using commercial PbO targets, (\lead \textsuperscript{16}O\textsubscript{2})\textsuperscript{-} currents are very stable, and typically in the range between 20 and 100~nA. When using processed PbO samples currents are stable as well, although with 5 times lower intensity. The lower currents in processed PbO targets show that the chemical methods which is used for the production of this material need to be improved. Therefore, a study of the chemical efficiency of the different steps for the production of PbO targets will be performed. Ion source output of \leadle\ from \fluoride\ targets is less stable, being really high during the first minute of sputtering, but declining exponentially during the early stage of the sputtering. After 20~min, ion current output stabilizes, but not to the same levels as in the case of PbO targets. Even using the same mix of commercial \fluoride\ with silver, a high variability is observed in the ion current output between targets, regardless whether the targets are commercial lead fluoride or processed samples. In some cases \leadle\ current output reaches several tens of nA; for others, after the first decay, it stays at levels closer to 5~nA.

Most efficiencies do not depend on the use of one or the other material. Transmission of the 3+ state in the Ar gas stripper is of 8\%, while the optical transmission through the post-accelerator spectrometer to the detector is 57\%. This optical transmission is estimated from the ratio between the directly measured \leadams/\lead\ ratio and the nominal \leadams/\lead\ ratio for reference samples. These reference samples are prepared by mixing:
\begin{itemize}
    \item $\sim$0.1~g of a solution with a \leadams\ concentration of 1.71~$\times$~10\textsuperscript{-16}~mol/g, prepared as a dilution of the NIST reference material SRM 4337 \cite{NIST-Pb210}.
    \item $\sim$0.27~g of a \leadams -free Pb carrier solution with a Pb concentration of 37.3~mg/g. This solution was prepared by dissolving a 1~cm\textsuperscript{3} cube of ancient lead provided by PNNL (the sample number 5 in Ref.~\cite{Orrell2016}) in diluted nitric acid. The final Pb concentration was measured by ICP-ES.
\end{itemize}

Therefore, the final \leadams/\lead\ ratio is around 1.5~$\times$~10\textsuperscript{-12}, slightly changing from one target to another depending on the exact masses of each solution. These masses are measured with a precision balance. \autoref{tab:optical-trans} shows the comparison between measured and nominal \leadams/\lead\ ratios for 3 reference targets produced with this method.

\begin{table}[t!]
    \caption{Parameters of \leadams\ measurements in the AEL-AMS system. \leadams /\lead\ blank ratio is presented in the next subsection.}
    \centering
    \begin{tabular}{||l||l|l||}
    \textbf{Target material} & \fluoride & PbO \\
    \textbf{Negative stable isotope ion} & (\textsuperscript{206}Pb\textsuperscript{19}F\textsubscript{3})\textsuperscript{-} & (\textsuperscript{206}Pb\textsuperscript{16}O\textsubscript{2})\textsuperscript{-}\\
    \textbf{Negative ion current output (nA)} & 5-20 & 20-100 \\
    \textbf{Charge state} & \multicolumn{2}{c||}{3+}\\
    \textbf{Terminal voltage (MV)} & \multicolumn{2}{c||}{2.5}\\
    \textbf{Stripper transmission (\%)} & \multicolumn{2}{c||}{8} \\
    \textbf{Accelerator to GIC optical transmission (\%)} & \multicolumn{2}{c||}{57}\\
    \end{tabular}   
    \label{tab:efficiency}
\end{table}

\begin{table}[h!]
    \caption{Experimental/nominal ratio from 3 reference targets. The average experimental/nominal ratio is of a 57~$\pm$~10~\%. This optical transmission is limited by the fact that in this work, the \leadamshe\ ions must pass through 3 additional tightly set slits before entering the final detector, whereas the \leadhe\ ions enter directly into an off-axis Faraday cup without loss. In future development, this parameter could be improved after systematic study of slits setting to balance background reduction and ion transmission.}
    \centering
    \begin{tabular}{lll}
    \hline 
    \hline
    \textbf{Nominal ratio ($\mathbf{\times}$ 10\textsuperscript{-12})} & \textbf{Measured ratio ($\mathbf{\times}$ 10\textsuperscript{-12})} & \textbf{Optical transmission (\%)} \\
    \hline
    \hline
    1.475 $\pm$ 0.016  & 0.78 $\pm$ 0.12 & 47.4 $\pm$ 8.5\\
    \hline
    1.543 $\pm$ 0.017 & 1.07 $\pm$ 0.15 & 64.0 $\pm$ 10 \\
    \hline
    1.534 $\pm$ 0.017 & 0.97 $\pm$ 0.16 & 58.4 $\pm$ 11 \\
    \hline
    \hline
    \end{tabular}   
    \label{tab:optical-trans}
\end{table}

\subsection{Background estimation}

One of the most important sources of background in \leadams\ AMS measurements is the interference of near isobaric \textsuperscript{208}Pb\textsuperscript{3+} ions. When using lead oxide samples, compared to lead fluoride, this interference is expected to be augmented by the injection of (\textsuperscript{208}Pb\textsuperscript{16}O\textsuperscript{18}O)\textsuperscript{-} ions with the (\textsuperscript{210}Pb\textsuperscript{16}O\textsubscript{2})\textsuperscript{-} ions (242~u) in the pre-accelerator spectrometer. Nevertheless, the post-accelerator spectrometer of the AEL-AMS system is designed to have a high mass resolution, which reduces the effect of this interference significantly.

In order to produce the processed blanks, the \leadams -free Pb carrier solution described previously is used. A comparison of \fluoride\ and PbO targets produced from this solution, with commercial ultra-pure \fluoride\ and PbO, respectively, allows us to assess the level of \leadams\ contamination during chemical processing of samples. It also let us determine if those commercial products can be considered for the blank material. Due to the relatively low ionization efficiencies, number of counts for blank samples are typically very low.

For \fluoride\ samples, the total number of counts from blank targets can be down to just 1 count. The blank \leadams /\lead\ ratio, before normalizing to the reference targets, is 3~$\times$~10\textsuperscript{-14}. This has a huge statistical uncertainty. No systematic difference has been observed between  \fluoride\ targets produced from the blank solution and commercial \fluoride , which shows that this material is clean enough to be considered blank material.

First tests of the \leadams /\lead\ ratio from oxide samples using commercial PbO suggested a background one order of magnitude higher than the one for fluoride samples. In later measurements a systematic difference is observed between the PbO targets produced from the blank solution and the commercial PbO targets. As an example, the results for the two kinds of target during our most recent measurement are presented in \autoref{tab:PbO-blank}. The lower current output does not explain the much lower number of counts for processed targets. Even when \leadams/\lead\ ratio for processed blanks includes a high uncertainty of a 97\% due to both standard deviation between targets and statistical uncertainty, there is still quite a significant difference when comparing with the \leadams/\lead\ from the commercial PbO. These results suggest that the background for oxide samples is not so much higher than the one for fluoride, and the commercial PbO material should not be considered a blank. 

\begin{table}[b!]
    \caption{\leadams /\lead\ ratio from 3 processed blank PbO targets and a commercial PbO target. Total measurement time was 7570 seconds per target.}
    \centering
    \begin{tabular}{llll}
    \hline 
    \hline
    \textbf{Target} & \textbf{\begin{tabular}{l}Number of\\ detector counts\end{tabular}} & \textbf{\begin{tabular}{l}Average \leadhe\\current (nA)\end{tabular}} & \textbf{\begin{tabular}{l}\leadams /\lead *\\($\mathbf{\times}$ 10\textsuperscript{-14})\end{tabular}}\\
    \hline
    \hline
    Commercial PbO & 29 & 4.91 & 37.5 $\pm$ 8.6\\
    \hline
    \begin{tabular}{l}Processed blank 1 \\Processed blank 2 \\Processed blank 3\end{tabular} & \begin{tabular}{l} 0 \\ 3 \\ 2 \end{tabular} & \begin{tabular}{l} 1.26 \\ 1.70 \\ 1.09 \end{tabular} & 7.8 $\pm$ 7.6\\
    \hline
    \hline
    \end{tabular}
    \begin{flushleft} \footnotesize{*) This ratio does not include the correction of optical transmission using reference samples. Actual background is, therefore, 4 times higher.} \end{flushleft}  
    \label{tab:PbO-blank}
\end{table}

\begin{table}[b!]
    \caption{\leadams\ specific activity related to the background of the measurements at the 3 MV AMS system compared with that of different radiometric techniques \cite{Ebaid} and ICP-MS \cite{LARIVIERE2005188}.}
    {\centering
    \begin{tabular}{llll}
    \hline
    \hline
    \textbf{Technique} & \begin{tabular}{l}\textbf{\leadams\ background}\\\textbf{(mBq)}\end{tabular} & \begin{tabular}{l}\textbf{Sample preparation}\\\textbf{time (days)}\end{tabular} & \begin{tabular}{l} \textbf{Measurement}\\\textbf{time (min)} \end{tabular} \\
    \hline
    \hline
    \leadams\ \textgamma -counting & 120 & 0 & 1000 \\
    \hline
    \textsuperscript{210}Bi \textbeta -counting & 20 & 10* & 1000 \\
    \hline
    \polonium\ \textalpha -counting & 0.4 & 90-180* & 1000 \\
    \hline
    ICP-MS & 90 & 4 & N/A \\
    \hline
    AMS (\fluoride) & $<$0.77 & 3 & 120 \\
    \hline
    AMS (PbO) & $<$1.9 & 3 & 120 \\    
    \hline
    \hline
    \end{tabular}}\vspace{6pt}\\
    \footnotesize{*) This sample preparation time includes the in-growth time required for the measured radionuclide to be in secular equilibrium with \leadams.}
    \label{tab:blk-techniques}
\end{table}

Including the correction to reference targets, the blank \leadams /\lead\ ratio for the fluoride targets is, in any case, lower than 1.1~$\times$~10\textsuperscript{-13}. Taking into account that 10~mg of Pb carrier per sample (2.41~mg of \lead ) are used, this \leadams /\lead\ ratio relates to a \leadams\ activity of 0.77~mBq. In the case of PbO targets, the corrected blank \leadams /\lead\ ratio is lower than 2.7~$\times$~10\textsuperscript{-13}, related to a \leadams\ activity of 1.9~mBq. In \autoref{tab:blk-techniques}, a comparison is presented of these levels with the backgrounds from other \leadams\ assay techniques. AMS clearly provides a lower background than ICP-MS and most radiometric techniques. In comparison with \polonium\ \textalpha-counting, the main advantage is that AMS requires a much shorter sample preparation time, because of the long time (3-6 months) required for \polonium\ to reach secular equilibrium with \leadams . In addition, \polonium\ \textalpha-counting is limited to the detection of the \leadams\ present at the surface of the material, while \leadams\ AMS gives information about the bulk of the material.

\section{Samples of interest for rare event searches}
\label{sec:application}

\subsection{Neutron production yield of \textalpha\ particles from the \textsuperscript{210}Pb chain}

The most problematic contribution of \leadams\ to the background in rare event searches is due to the \textalpha\ decay of its daughter, \polonium . This is mainly because of the (\textalpha ,n) reactions which may happen when low-Z elements are present in the detector materials. \textalpha\ particles from \polonium\ decay have, typically, a range of just several tens of \textmu m in most solid materials. Therefore, (\textalpha,n) reactions typically take place in the material where the \polonium\ decay occurred.

\begin{table}[b!]
\caption{Neutron yields of the \textalpha\ particles from the \leadams\ chain in different materials, obtained with the NeuCBOT tool \cite{Westerdale2017}.
\vspace{7pt}}
\centering
    \begin{tabular}{lll}
    \hline
    \hline
   \textbf{Material} & \textbf{Formula} & \textbf{Neutron yield (n/decay)} \\
   \hline
   \hline
   Acrylic (PMMA) & [C\textsubscript{5}O\textsubscript{2}H\textsubscript{8}]\textsubscript{n} & 1.16 $\times$ 10\textsuperscript{-7}\\
   \hline
   Aluminum & Al & 1.49 $\times$ 10\textsuperscript{-6}\\
   \hline
   Kapton polyimide & [C\textsubscript{22}O\textsubscript{5}N\textsubscript{2}H\textsubscript{10}]\textsubscript{n} & 8.37 $\times$ 10\textsuperscript{-8}\\
   \hline
   Polyethylene & [C\textsubscript{2}H\textsubscript{4}]\textsubscript{n} & 1.30 $\times$ 10\textsuperscript{-7}\\
   \hline
   PVC & [C\textsubscript{2}H\textsubscript{3}Cl]\textsubscript{n} & 7.38  $\times$ 10\textsuperscript{-8}\\
   \hline
   Quartz & SiO\textsubscript{2} & 7.91 $\times$ 10\textsuperscript{-8}\\
   \hline
   Sapphire & Al\textsubscript{2}O\textsubscript{3} &  7.35 $\times$ 10\textsuperscript{-7}\\
   \hline
   Silicon & Si & 1.17 $\times$ 10\textsuperscript{-7} \\
   \hline
   Silicon carbide & SiC & 1.45 $\times$ 10\textsuperscript{-7} \\
   \hline
   Teflon (PTFE/FEP) & [C\textsubscript{2}F\textsubscript{4}]\textsubscript{n}/[C\textsubscript{5}F\textsubscript{10}]\textsubscript{n} & 9.37 $\times$ 10\textsuperscript{-6}\\
   \hline
   Titanium & Ti & 1.17 $\times$ 10\textsuperscript{-8} \\
    \hline
    \hline
    \end{tabular}
    \label{tab:neucbot}
\end{table}

Consequently, the impact on the final background of the experiment by the \leadams\ concentrations will depend on the material in which this \leadams\ is present. The neutron yield of the \textalpha\ particles from the \leadams\ chain (mainly from \polonium\ decay) in different materials was studied with the NeuCBOT tool \cite{Westerdale2017}. In \autoref{tab:neucbot}, some examples of these yields in materials commonly used in rare event search experiments are presented. Focusing on the different kinds of polymer materials, a huge difference is observed between fluorinated ones, like Teflon, and others. In the case of Kapton and acrylic, neutron yields are around 10\textsuperscript{-7} neutrons per decay, while this yield is almost 100 times higher for Teflon. Therefore, \leadams\ concentrations in materials like Kapton FN, consisting on Kapton polyimide coated with Teflon FEP, will have a much higher impact than equal \leadams\ concentrations in other kind of polyimide.


\subsection{Measurement of polymer samples}

\begin{wrapfigure}{L}{0.33\textwidth}
    \centering

\includegraphics[width=0.31\textwidth]{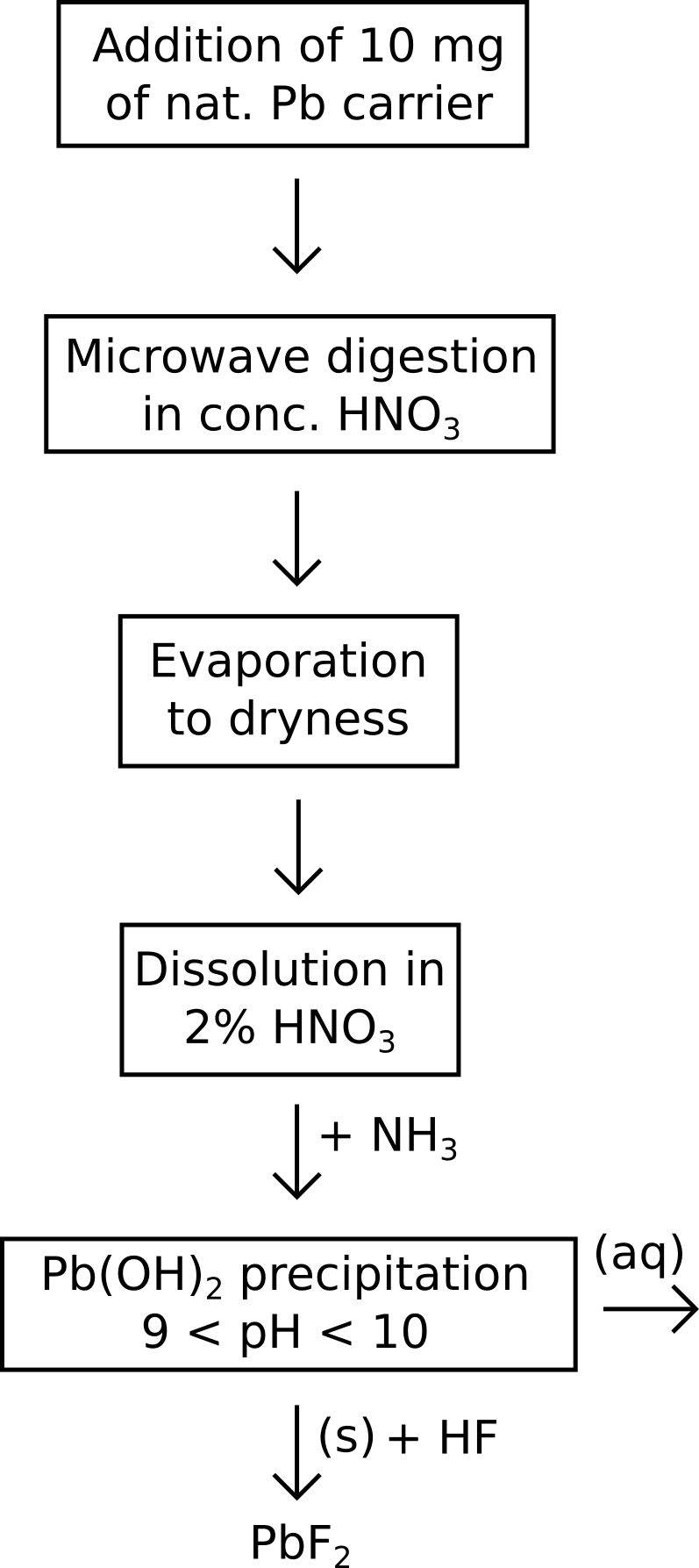}

    \caption{Flowchart of the basic chemical steps for the preparation to lead fluoride targets for Kapton samples. Note that, following microwave digestion, all processes conserve the \leadams /\lead\ ratio.}
    \label{fig:leadams-flowchart}
\end{wrapfigure}

The first type of sample assayed is ultra-thin Kapton film. Kapton is expected to be largely used as electric insulator in circuits used in nEXO. Therefore, it is one of the materials whose radio-characterization is of most interest in the nEXO collaboration \cite{nEXO2018-sensitivity}.

A chemical procedure to completely digest Kapton, extracting \leadams, and precipitate lead fluoride samples was developed. A flowchart of this procedure is shown in \autoref{fig:leadams-flowchart}. Most of the procedure can be applied to other polymers, especially those where only C, H and O are present. A pressurized microwave digestion system (CEM MARS 230/60) is used along with concentrated HNO\textsubscript{3}, as suggested by the bibliography \cite{ARNQUIST2020163573}.

First, the sample is cleaned with 7~mol·L\textsuperscript{-1}~HNO\textsubscript{3}. After discarding the cleaning solution, concentrated HNO\textsubscript{3} (70\% v/v) is added, together with the natural Pb carrier.

The microwave digestion procedure for Kapton is the following:

\begin{itemize}
    \item[1.] Initial ramp up to 150~ºC in 10~min.
    \item[2.] Hold at 150~ºC for 10 min.
    \item[3.] Ramp up up to 240~ºC in 10~min
    \item[4.] Hold at 240~ºC for 15 min.
    \item[5.] Cool down for 15~min.
\end{itemize}

During each operation of the microwave digestion system, less than 0.5~g of organic material can be digested. Therefore, in order to be able to digest up to 1.5-2.0~g of polymer sample, the 10~mg Pb carrier can be dissolved in 40~mL of concentrated HNO\textsubscript{3}. This way, separating the total sample of polymer in 4 aliquots of $<$0.5~g, performing 4 replicates with 10~mL of the concentrated HNO\textsubscript{3}~+~Pb~carrier solution (one per aliquot), and finally mixing of the 4 solutions, allow us to digest such a relatively large amount of sample. The digestion is tested by centrifuging the final solution, and checking that no precipitate is formed.  It is assumed, so, that at least, a 99\%  of the sample is completely dissolved. The subsequent 1\% of uncertainty will be negligible in comparison with the high statistic uncertainty due to the low number of detected counts. The \leadams /\lead\ ratio in this solution is the same one as in the final \fluoride\ precipitate, establishing a direct relationship between this ratio and the \leadams\ concentration in the original Kapton sample.

This solution is then fully evaporated, leaving the solid lead nitrate, together with residual organic material. 2\%~v/v~HNO\textsubscript{3} is added and the Pb(OH)\textsubscript{2} is precipitated by adding NH\textsubscript{3}. Samples are left to settle overnight, and centrifuged, discarding the supernatant. Fluoride is formed by acid-base reaction of the \hydroxide\ precipitate with HF. In this last step, residual organic material remains in the HF solution, so the final target is free of this residue. Targets produced by this method presented similar current outputs as those from commercial \fluoride .

In order to produce lead oxide, instead of fluoride samples, the precipitation of lead hydroxide is replaced by lead carbonate precipitation. This is performed by adding saturated sodium carbonate aqueous solution, instead of ammonia. After centrifuging and discarding the supernatant, the carbonate precipitate is heated in a muffle furnace for 6~h at a temperature of 650~ºC, producing PbO by thermal decomposition. 

Initial measurements of the Kapton samples, using lead fluoride targets, are presented in \autoref{tab:210pb-results-polymers}. During these measurements, all the samples gave \leadams/\lead\ ratios lower or equal to blank levels. Therefore, only 90\% C.L. upper limits of the \leadams\ specific activity can be given. These upper limits are calculated using the method of Feldman and Cousins \cite{Feldman1998}, taking into account the high uncertainty in the background level, i.e., using the lower limit of the background to obtain the most conservative upper limit for the \leadams\ specific activity. With these results, it can stated that the maximum \leadams\ specific activity in the 100HN Kapton material is 1.1~Bq/kg. Taking into account that this film has a thickness of 25.4~\textmu m, and that its density is 1.42~g·cm\textsuperscript{-3}, the upper limit for the \leadams\ activity per unit of surface is 40~mBq·m\textsuperscript{-2}. To provide an idea of the advantage of this level of measurement, the \leadams\ plating out rate in high-density polyethylene in the underground lab of SNOLAB, due to the \textsuperscript{222}Rn concentration there, is 2.46~mBq·m\textsuperscript{-2}·d\textsuperscript{-1}~\cite{STEIN201892}.

This is already very useful information for the nEXO experiment to calculate the maximum impact to background that this \leadams\ concentration in Kapton can have. Besides, it compliments ICP-MS measurements of \textsuperscript{238}U in this same material performed at PNNL. This \textsuperscript{238}U specific activity is 0.012~Bq/kg \cite{ARNQUIST2020163573}. Assuming that, at a minimum, \leadams\ specific activity should be in secular equilibrium with its parent \textsuperscript{238}U, this specific activity would be in the 0.012-1.1~mBq/kg range.

In order to test the chemical methods for PbO, measurements of different Kapton samples using oxide as target material were also performed. Results, which are shown in \autoref{tab:210pb-results-polymers-PbO}, allow only to establish 90\% C.L. upper limits for the \leadams\ in these samples.

\begin{table}[h!]
\caption{\leadams\ concentrations measured by AMS in different aliquots of a sample of 100HN~Kapton film, using lead fluoride targets.}
\centering
{\small \begin{tabular}{lllll}
    \hline
    \hline
    \textbf{Aliquot} & \textbf{\begin{tabular}{l}Sample\\mass (g)\end{tabular}} & \textbf{\begin{tabular}{l}\lead\ carrier\\(mg)\end{tabular}} & \textbf{\begin{tabular}{l}\leadams /\lead \\($\times$ 10\textsuperscript{-13})\end{tabular}} & \textbf{\begin{tabular}{l}[\leadams ] \\(Bq/kg)\end{tabular}}\\
    \hline
    \hline
    Pb-Kapton-191025 & 1.3763 $\pm$ 0.0020 & 2.4749 $\pm$ 0.0090 & $<$ 1.4 & $<$ 0.74 \\
    Pb-Kapton-191107 & 1.5995 $\pm$ 0.0020 & 2.4064 $\pm$ 0.0090 & $<$ 2.3 & $<$ 1.0\\
    Pb-Kapton-200114 & 1.6157 $\pm$ 0.0020 & 2.4488 $\pm$ 0.0090 & $<$ 2.5 & $<$ 1.1\\
    \hline
    \hline
    \end{tabular}}
    \label{tab:210pb-results-polymers}
\end{table}

\begin{table}[h!]
\caption{\leadams\ concentrations measured by AMS in different Kapton HN film samples, using lead oxide targets. 
}
\centering
{\small \begin{tabular}{lllll}
    \hline
    \hline
    \textbf{Sample}  & \textbf{\begin{tabular}{l}Sample\\mass (g)\end{tabular}} & \textbf{\begin{tabular}{l}\lead\ carrier\\(mg)\end{tabular}} & \textbf{\begin{tabular}{l}\leadams /\lead \\($\times$ 10\textsuperscript{-13})\end{tabular} }& \textbf{\begin{tabular}{l}[\leadams ] \\(Bq/kg)\end{tabular}}\\
    \hline
    \hline
    100HN   & 1.4692 $\pm$ 0.0020  & 2.4317 $\pm$ 0.0090 & $<$ 3.1 & $<$ 1.5 \\
    200HN   & 1.6131 $\pm$ 0.0020  & 2.4551 $\pm$ 0.0090 & $<$ 4.8 & $<$ 2.1 \\
    300HN   & 1.5985 $\pm$ 0.0020  & 2.4263 $\pm$ 0.0090 & $<$ 6.3 & $<$ 2.8 \\
    500HN  & 1.7928 $\pm$ 0.0020  & 2.4632 $\pm$ 0.0090 & $<$ 2.6 & $<$ 1.0 \\
    \hline
    \hline
    \end{tabular}}
    \label{tab:210pb-results-polymers-PbO}
\end{table}

\section{Conclusions and prospects}
\label{sec:conclusion}

Even though the \leadams\ half-life is much shorter than those of radionuclides commonly measured by AMS, this technique still provides a much better sensitivity than \textgamma -counting. In the 3 MV AMS system at the University of Ottawa, \leadams /\lead\ background is associated to a \leadams\ activity close to 1~mBq. In comparison with \polonium\ \textalpha -counting, which requires 3-6 months after chemical treatment to achieve the secular equilibrium, AMS provides much faster results and provides information about the whole sample, and not only the surface activity. 

A study of the ion source current output of fluoride targets as a function of the temperature of the Cs reservoir will be performed to assess whether it is possible to improve it without increasing the instability. Furthermore, the optical transmission for measurements with fluoride targets could be improved using a less conservative set-up of the slits, but ensuring that the background is not severely increased.

A better assay of the background for measurements with lead oxide samples is required. Preliminary results suggest that this background may not be much higher than the one observed with fluoride targets. If this is confirmed, it shows the very good mass resolution of the post-accelerator spectrometer of this AMS system for suppressing the scatter tails of the nearby \textsuperscript{208}Pb\textsuperscript{3+} ions.

In any case, with the current performance parameters, \leadams\ measurements at the 3~MV AMS system at the University of Ottawa have already the potential of performing groundbreaking assay of materials for Astroparticle Physics experiments. Besides recent work for the nEXO collaboration, experiments involved in direct detection of dark matter have already expressed their interest in the capabilities of \leadams\ measurements at this AMS facility.

\section*{Acknowledgments}

The authors are deeply indebted to the André E. Lalonde Laboratory, at the University of Ottawa, for the access to the 3 MV AMS system to perform the measurements, to their Actinides and Fission Products Laboratory for the preparation of the samples; and to the Geochemistry Laboratory at the University of Ottawa, for the access to their microwave digestion system and for the ICP-ES measurement of the Pb concentration of the blank solution. We would like to thank Dr. Shawn Westerdale for his help and updates of the NeuCBOT tool, used to calculate the neutron production yield of the \textalpha\ particles from the \leadams\ chain. This work has been supported by the Arthur B. McDonald Canadian Astroparticle Physics Research Institute, the National Research Council of Canada  and the Canada Foundation for Innovation. Measurements in this work were performed for the nEXO collaboration, and the authors are thankful to all the members of the collaboration for their advice, motivation and support.

\bibliographystyle{elsarticle-num}
\bibliography{library}

\begin{thebibliography}{10}
\expandafter\ifx\csname url\endcsname\relax
  \def\url#1{\texttt{#1}}\fi
\expandafter\ifx\csname urlprefix\endcsname\relax\def\urlprefix{URL }\fi
\expandafter\ifx\csname href\endcsname\relax
  \def\href#1#2{#2} \def\path#1{#1}\fi

\bibitem{NEWSG}
{NEWS-G collaboration}, {First results from the NEWS-G direct dark matter
  search experiment at the LSM}, Astroparticle Physics 97 (2018) 54 -- 62.
\newblock \href {http://arxiv.org/abs/1706.04934} {\path{arXiv:1706.04934}},
  \href {https://doi.org/https://doi.org/10.1016/j.astropartphys.2017.10.009}
  {\path{doi:https://doi.org/10.1016/j.astropartphys.2017.10.009}}.

\bibitem{PICO}
{PICO collaboration}, {Dark matter search results from the complete exposure of
  the PICO-60 ${\mathrm{C}}_{3}{\mathrm{F}}_{8}$ bubble chamber}, Phys. Rev. D
  100 (2019) 022001.
\newblock \href {http://arxiv.org/abs/1902.04031} {\path{arXiv:1902.04031}},
  \href {https://doi.org/10.1103/PhysRevD.100.022001}
  {\path{doi:10.1103/PhysRevD.100.022001}}.

\bibitem{DEAP2018}
{DEAP collaboration}, {Design and construction of the DEAP-3600 dark matter
  detector}, Astroparticle Physics 108 (2019) 1 -- 23.
\newblock \href {http://arxiv.org/abs/1712.01982} {\path{arXiv:1712.01982}},
  \href {https://doi.org/10.1016/j.astropartphys.2018.09.006}
  {\path{doi:10.1016/j.astropartphys.2018.09.006}}.

\bibitem{Lehnert2018}
B.~Lehnert, {Backgrounds in the DEAP-3600 Dark Matter Experiment} (2018).
\newblock \href {http://arxiv.org/abs/1805.06073} {\path{arXiv:1805.06073}}.

\bibitem{EXO-200}
{EXO-200 collaboration}, {Search for Neutrinoless Double-$\ensuremath{\beta}$
  Decay with the Complete EXO-200 Dataset}, Phys. Rev. Lett. 123 (2019) 161802.
\newblock \href {http://arxiv.org/abs/1906.02723} {\path{arXiv:1906.02723}},
  \href {https://doi.org/10.1103/PhysRevLett.123.161802}
  {\path{doi:10.1103/PhysRevLett.123.161802}}.

\bibitem{EXO-radioassay}
D.~Leonard, D.~Auty, T.~Didberidze, R.~Gornea, P.~Grinberg, R.~MacLellan,
  B.~Methven, A.~Piepke, J.-L. Vuilleumier, {EXO-200 collaboration}, {Trace
  radioactive impurities in final construction materials for EXO-200}, Nuclear
  Instruments and Methods in Physics Research Section A: Accelerators,
  Spectrometers, Detectors and Associated Equipment 871 (2017) 169 -- 179.
\newblock \href {http://arxiv.org/abs/1703.10799} {\path{arXiv:1703.10799}},
  \href {https://doi.org/10.1016/j.nima.2017.04.049}
  {\path{doi:10.1016/j.nima.2017.04.049}}.

\bibitem{Majorana2019}
{Majorana Collaboration}, {Search for neutrinoless double-$\ensuremath{\beta}$
  decay in $^{76}\mathrm{Ge}$ with 26 kg yr of exposure from the Majorana
  Demonstrator}, Phys. Rev. C 100 (2019) 025501.
\newblock \href {http://arxiv.org/abs/1902.02299} {\path{arXiv:1902.02299}},
  \href {https://doi.org/10.1103/PhysRevC.100.025501}
  {\path{doi:10.1103/PhysRevC.100.025501}}.

\bibitem{Majorana-radioassay}
{Majorana collaboration}, The majorana demonstrator radioassay program, Nuclear
  Instruments and Methods in Physics Research Section A: Accelerators,
  Spectrometers, Detectors and Associated Equipment 828 (2016) 22 -- 36.
\newblock \href {http://arxiv.org/abs/1601.03779} {\path{arXiv:1601.03779}},
  \href {https://doi.org/https://doi.org/10.1016/j.nima.2016.04.070}
  {\path{doi:https://doi.org/10.1016/j.nima.2016.04.070}}.

\bibitem{nEXO-PCR}
{nEXO collaboration}, {nEXO Pre-Conceptual Design Report} (2018).
\newblock \href {http://arxiv.org/abs/1805.11142} {\path{arXiv:1805.11142}}.

\bibitem{nEXO2018-sensitivity}
{nEXO collaboration}, {Sensitivity and discovery potential of the proposed nEXO
  experiment to neutrinoless double-\textbeta\ decay}, Physical Review C 97
  (2018) 065503.
\newblock \href {http://arxiv.org/abs/1710.05075} {\path{arXiv:1710.05075}},
  \href {https://doi.org/10.1103/PhysRevC.97.065503}
  {\path{doi:10.1103/PhysRevC.97.065503}}.

\bibitem{STEIN201892}
M.~Stein, D.~Bauer, R.~Bunker, R.~Calkins, J.~Cooley, B.~Loer, S.~Scorza,
  {Radon daughter plate-out measurements at SNOLAB for polyethylene and
  copper}, Nuclear Instruments and Methods in Physics Research Section A:
  Accelerators, Spectrometers, Detectors and Associated Equipment 880 (2018) 92
  -- 97.
\newblock \href {http://arxiv.org/abs/1708.09476} {\path{arXiv:1708.09476}},
  \href {https://doi.org/10.1016/j.nima.2017.10.054}
  {\path{doi:10.1016/j.nima.2017.10.054}}.

\bibitem{HOU2008105}
X.~Hou, P.~Roos, Critical comparison of radiometric and mass spectrometric
  methods for the determination of radionuclides in environmental, biological
  and nuclear waste samples, Analytica Chimica Acta 608 (2008) 105 -- 139.
\newblock \href {https://doi.org/10.1016/j.aca.2007.12.012}
  {\path{doi:10.1016/j.aca.2007.12.012}}.

\bibitem{Ebaid}
Y.~Ebaid, A.~Khater, {Determination of \textsuperscript{210}Pb in environmental
  samples}, Journal of Radioanalytical and Nuclear Chemistry 270 (2006) 609 --
  619.
\newblock \href {https://doi.org/10.1007/s10967-006-0470-5}
  {\path{doi:10.1007/s10967-006-0470-5}}.

\bibitem{LARIVIERE2005188}
D.~Larivière, K.~Reiber, R.~Evans, R.~Cornett, {Determination of 210Pb at
  ultra-trace levels in water by ICP-MS}, Analytica Chimica Acta 549 (2005) 188
  -- 196.
\newblock \href {https://doi.org/10.1016/j.aca.2005.06.020}
  {\path{doi:10.1016/j.aca.2005.06.020}}.

\bibitem{Steier2002}
P.~Steier, R.~Golser, W.~Kutschera, V.~Liechtenstein, A.~Priller, A.~Valenta,
  C.~Vockenhuber, {Heavy ion AMS with a “small” accelerator"}, Nuclear
  Instruments and Methods in Physics Research Section B: Beam Interactions with
  Materials and Atoms 188 (2002) 283 -- 287.
\newblock \href {https://doi.org/10.1016/S0168-583X(01)01114-4}
  {\path{doi:10.1016/S0168-583X(01)01114-4}}.

\bibitem{VOCKENHUBER2003713}
C.~Vockenhuber, I.~Ahmad, R.~Golser, W.~Kutschera, V.~Liechtenstein,
  A.~Priller, P.~Steier, S.~Winkler, Accelerator mass spectrometry of heavy
  long-lived radionuclides, International Journal of Mass Spectrometry 223-224
  (2003) 713 -- 732.
\newblock \href {https://doi.org/10.1016/S1387-3806(02)00944-2}
  {\path{doi:10.1016/S1387-3806(02)00944-2}}.

\bibitem{Sookdeo2015}
A.~Sookdeo, J.~Cornett, W.~E. Kieser, {Optimizing production of Pb beams for
  \textsuperscript{205,210}Pb analysis by Accelerator Mass Spectrometry},
  Nuclear Instruments and Methods in Physics Research Section B: Beam
  Interactions with Materials and Atoms 361 (2015) 450 -- 453.
\newblock \href {https://doi.org/10.1016/j.nimb.2015.02.063}
  {\path{doi:10.1016/j.nimb.2015.02.063}}.

\bibitem{Sookdeo2016}
A.~Sookdeo, R.~J. Cornett, X.-L. Zhao, C.~R.~J. Charles, W.~E. Kieser,
  {Measuring \textsuperscript{210}Pb by accelerator mass spectrometry: a study
  of isobaric interferences of \textsuperscript{204,205,208}Pb and
  \textsuperscript{210}Pb}, Rapid Communications in Mass Spectrometry 30 (2016)
  867--872.
\newblock \href {https://doi.org/10.1002/rcm.7501}
  {\path{doi:10.1002/rcm.7501}}.

\bibitem{KORSCHINEK1988328}
G.~Korschinek, J.~Sellmair, A.~Urban, M.~Müller, {A study of different ion
  sources for use in the \textsuperscript{205}Pb experiment}, Nuclear
  Instruments and Methods in Physics Research Section A: Accelerators,
  Spectrometers, Detectors and Associated Equipment 271 (1988) 328 -- 331.
\newblock \href {https://doi.org/10.1016/0168-9002(88)90180-5}
  {\path{doi:10.1016/0168-9002(88)90180-5}}.

\bibitem{Zhao2010}
X.-L. Zhao, A.~Litherland, J.~Eliades, W.~Kieser, Q.~Liu, {Studies of anions
  from sputtering I: Survey of MF\textsubscript{n}\textsuperscript{-}}, Nuclear
  Instruments and Methods in Physics Research Section B: Beam Interactions with
  Materials and Atoms 268 (2010) 807 -- 811.
\newblock \href {https://doi.org/10.1016/j.nimb.2009.10.036}
  {\path{doi:10.1016/j.nimb.2009.10.036}}.

\bibitem{cookbook}
R.~Middleton, \href{http://www.pelletron.com/cookbook.pdf}{{A Negative-Ion
  Cookbook}}, Department Of Physics, University of Pennsylvania, 1990.
\newline\urlprefix\url{http://www.pelletron.com/cookbook.pdf}

\bibitem{Kieser2015}
W.~Kieser, X.-L. Zhao, I.~Clark, R.~Cornett, A.~Litherland, M.~Klein, D.~Mous,
  J.-F. Alary, {The André E. Lalonde AMS Laboratory – The new accelerator
  mass spectrometry facility at the University of Ottawa}, Nuclear Instruments
  and Methods in Physics Research Section B: Beam Interactions with Materials
  and Atoms 361 (2015) 110 -- 114.
\newblock \href {https://doi.org/10.1016/j.nimb.2015.03.014}
  {\path{doi:10.1016/j.nimb.2015.03.014}}.

\bibitem{NIST-Pb210}
{National Institute of Standards and Technology}, {SRM 4337 - Lead-210
  Radioactivity Standard},
  \url{https://www-s.nist.gov/srmors/view_detail.cfm?srm=4337}.

\bibitem{Orrell2016}
J.~L. Orrell, C.~E. Aalseth, I.~J. Arnquist, T.~A. Eggemeyer, B.~D. Glasgow,
  E.~W. Hoppe, M.~E. Keillor, S.~M. Morley, A.~W. Myers, C.~T. Overman, S.~M.
  Shaff, K.~S. Thommasson, {Assay methods for 238U, 232Th, and 210Pb in lead
  and calibration of 210Bi Bremsstrahlung emission from lead}, Journal of
  Radioanalytical and Nuclear Chemistry 309 (2016) 1271--1281.
\newblock \href {https://doi.org/10.1007/s10967-016-4732-6}
  {\path{doi:10.1007/s10967-016-4732-6}}.

\bibitem{Westerdale2017}
S.~Westerdale, P.~Meyers, Radiogenic neutron yield calculations for
  low-background experiments, Nuclear Instruments and Methods in Physics
  Research Section A: Accelerators, Spectrometers, Detectors and Associated
  Equipment 875 (2017) 57 -- 64.
\newblock \href {http://arxiv.org/abs/1702.02465} {\path{arXiv:1702.02465}},
  \href {https://doi.org/10.1016/j.nima.2017.09.007}
  {\path{doi:10.1016/j.nima.2017.09.007}}.

\bibitem{ARNQUIST2020163573}
I.~J. Arnquist, C.~Beck, M.~L. di~Vacri, K.~Harouaka, R.~Saldanha, {Ultra-low
  radioactivity Kapton and copper-Kapton laminates}, Nuclear Instruments and
  Methods in Physics Research Section A: Accelerators, Spectrometers, Detectors
  and Associated Equipment 959 (2020) 163573.
\newblock \href {https://doi.org/10.1016/j.nima.2020.163573}
  {\path{doi:10.1016/j.nima.2020.163573}}.

\bibitem{Feldman1998}
G.~J. Feldman, R.~D. Cousins, {Unified approach to the classical statistical
  analysis of small signals}, Phys. Rev. D 57 (1998) 3873--3889.
\newblock \href {http://arxiv.org/abs/physics/9711021}
  {\path{arXiv:physics/9711021}}, \href
  {https://doi.org/10.1103/PhysRevD.57.3873}
  {\path{doi:10.1103/PhysRevD.57.3873}}.

\end{thebibliography}

\end{document}